# Three-Phase Dynamic Simulation of Power Systems Using Combined Transmission and Distribution System Models

Himanshu Jain[1], Abhineet Parchure[1], Robert P. Broadwater[1], *Member, IEEE*, Murat Dilek[2], and Jeremy Woyak[2]

*Abstract*-- **This paper presents a new method for studying electromechanical transients in power systems using three phase, combined transmission and distribution models (hybrid models). The methodology models individual phases of an electric network and associated unbalance in load and generation. Therefore, the impacts of load unbalance, single phase distributed generation and line impedance unbalance on electromechanical transients can be studied without using electromagnetic transient simulation (EMTP) programs. The implementation of this methodology in software is called the Three Phase Dynamics Analyzer (TPDA). Case studies included in the paper demonstrate the accuracy of TPDA and its ability to simulate electromechanical transients in hybrid models. TPDA has the potential for providing electric utilities and power system planners with more accurate assessment of system stability than traditional dynamic simulation software that assume balanced network topology.**

*Index Terms*— **EMTP, Power System Stability, Distributed Power Generation, Power Quality, Power System Restoration**

## I. INTRODUCTION

IN recent years solar photovoltaic (hereafter referred to as PV) based Distributed Generation (DG) has become very popular. The reference case in the 2014 Annual Energy Outlook released by the Energy Information Administration (EIA) highlights this popularity, as it assumes that PV and wind will continue to dominate the new commercial DG capacity and account for 62.3% of the total commercial DG capacity in 2040, thereby providing about 10% of the total electric energy generated in the US [1]. However, the distribution of DG across the nation is not going to be uniform. Some regions and utilities will have more DG than others. For example, the state of California is a leader in PV generation with the installed capacity of rooftop solar PV at 2 GW in 2013 [2]. Utilities in regions with large numbers of DGs connected to their network are already facing integration challenges, which are only going to increase in the future. To effectively address the challenge of integrating DG and ensuring the reliability of the electric grid, planners and operators need new modeling and analysis tools that can provide them with accurate information about the impact of adding DG, and also to help them formulate strategies to mitigate adverse impacts. With this objective in mind, we present in this paper a new approach for studying electromechanical transients in power systems using three phase hybrid models to facilitate a more comprehensive investigation of the impacts of adding DG in the electric grid. The software implementation of this methodology is called the Three Phase Dynamics Analyzer, or TPDA.

So far many authors have attempted to develop approaches for studying the dynamics of distribution systems, particularly under different penetration levels of DG. Many of these studies have used balanced network representations of distribution systems [3], [4] and/or small networks to test their approaches [3]-[7]. The authors in [8] developed models for synchronous DGs and Doubly Fed Induction Generators (DFIGs), and used PSCAD/EMTPDC software to perform transient stability studies on unbalanced networks with DGs. Authors in [9] developed a method for performing three phase power flow analysis on an unbalanced network and used the results of the power flow to obtain an equivalent positive sequence network that included contributions from the negative and zero sequence networks. The equivalent positive sequence network was used for performing dynamic simulations. It was not clear how the authors could completely incorporate the impact of negative and zero sequence networks in the equivalent positive sequence network, since these networks cannot be decoupled under general unbalance (unbalance at more than one location in an electric network). Authors in [10] studied the impact of DG on the bulk transmission system, but used a balanced network for their analysis. This paper was interesting because it highlighted the need for studying the impacts of DG connected to the distribution network on bulk transmission, sentiments echoed by utility engineers as mentioned in a 2013 California Public Utilities Commission report [11]. An interesting mathematical model to study the small signal stability of unbalanced power systems was presented in [12]. The work of the authors in [12] was unique because, unlike in a balanced power system, a static equilibrium point for linearization cannot be defined in unbalanced systems. However, to develop the model the authors assumed that the synchronous machine dynamics can be "*separated into its respective sequence components*". Since the sequence components cannot be decoupled under general unbalance in a network, we are of the view that more experiments need to be conducted to identify the degree of unbalance up to which the assumptions of the paper can be justified.

---

[1]Himanshu Jain, Abhineet Parchure and Robert Broadwater are with the Electrical and Computer Engineering Department at Virginia Tech, Blacksburg, VA 24061 (e-mail: jhim86@vt.edu;abhineet@vt.edu;dew@vt.edu).
[2] Murat Dilek and Jeremy Woyak are with Electrical Distribution Design (EDD), Blacksburg, VA 24060,USA (e-mail: murat-dilek@edd-us.com; jeremy.woyak@gmail.com)

Based on the above discussion it may be concluded that studying electromechanical transients in unbalanced networks is a difficult problem, and short of modeling the electric network using differential equations in EMTP programs, simplifying assumptions must be made to make the problem tractable. Moreover, we did not come across a study that simulated power system dynamics using hybrid models.

In light of these observations, this paper presents an algorithm that enables the study of electromechanical transients in unbalanced networks without using EMTP programs and without assuming the network to be balanced, an assumption commonly made in commercial electromechanical transient simulation software.

## II. Concepts and Algorithm Behind TPDA

Study of electromechanical transients in power systems involves the formulation and solution of a set of Differential Algebraic Equations (DAEs) [13], [14]. Commercial software that are used for studying electromechanical transients, such as GE-PSLF® and PTI-PSS/E®, assume the network to be balanced. Under this assumption the solution of DAEs is simplified because Park's transformation enables direct conversion from $dq0$ frame voltages and currents in the time domain to corresponding phasors in the frequency domain [13]. The DAEs formed using unbalanced three phase network models, however, do not offer such simplification because six unknown quantities (phasor magnitudes and angles of the three phases) at an instant need to be estimated in the $abc$ reference frame from three instantaneous $dq0$ reference frame quantities. Discussion about a mathematically sound solution of this problem is the primary focus of this section. This solution is implemented in TPDA and distinguishes it from existing electromechanical transient simulation software. Reference [15] was the only reference found that presented a method for obtaining the three phase phasors from $dq0$ frame quantities. However, the justification for the formula used was not presented.

The discussion that follows assumes for simplicity that synchronous generators are the only active devices in the network that act as voltage sources. However, TPDA can include any active device that can be modeled in three phase.

### A. Calculation of Six $dq0$ frame Voltages

The first step for obtaining three unique voltage phasors is to calculate six $dq0$ frame voltages. Generator stator algebraic equations (1)-(3) are used to obtain these voltages. The interested reader is referred to [13], [14] and Manuals of GE-PSLF® and PTI-PSS/E® for a detailed discussion and derivation of synchronous generator equations. Table I describes the symbols used in (1)-(3).

$$v_d = -R_s I_d - \frac{\omega}{\omega_s}\psi_q \quad (1)$$

$$v_q = -R_s I_q + \frac{\omega}{\omega_s}\psi_d \quad (2)$$

$$v_0 = -R_s I_0 \quad (3)$$

TABLE I
Definition of Symbols Used in (1)-(3)

| Symbol | Definition | Symbol | Definition |
|---|---|---|---|
| $\psi_d, \psi_q$ | $d, q$ axis fluxes | $\omega_s$ | Synchronous Speed (radians/second) |
| $v_d, v_q, v_0$ | $d, q, 0$ axis voltages | $R_s$ | Stator resistance |
| $I_d, I_q, I_0$ | $d, q, 0$ axis currents | $\omega$ | Rotor Speed (also a dynamic state variable); |

Before proceeding further it is important to mention that similar to the commercial software that use balanced network models for studying electromechanical transients, TPDA assumes that the network frequency stays fixed at 60 Hz. For studying electromechanical transients this assumption introduces negligible error in the simulation results as system frequency deviates little from 60 Hz [14]. This assumption, along with the modeling of the electric network as algebraic equations, allows network equations to be solved using a nonlinear equations solver (e.g., a modified power flow analysis program).

Calculation of six $dq0$ frame voltages in TPDA at every simulation iteration is now discussed. Let us assume that the current simulation time instant is $t$ and the updated voltage phasor is to be obtained for time $t + \Delta T$. TPDA first solves the network algebraic equations at time $t$ using the voltage phasors at generator terminals $V_a e^{j\beta_a}, V_b e^{j\beta_b}, V_c e^{j\beta_c}$ to obtain new current phasors $I_a e^{j\gamma_a}, I_b e^{j\gamma_b}, I_c e^{j\gamma_c}$. These current phasors are converted into instantaneous currents $i_a(t), i_b(t), i_c(t)$ using (4)-(6). Park's transform is used to calculate $dq0$ frame currents $\boldsymbol{I}_{dq0}(t)$ from $\boldsymbol{i}_{abc}(t)$, where $\boldsymbol{i}_{abc}(t) = [i_a(t)\ i_b(t)\ i_c(t)]^{\mathrm{T}}$.

$$i_a(t) = \sqrt{2}I_a \cos(2\pi * 60 * t + \gamma_a) \quad (4)$$

$$i_b(t) = \sqrt{2}I_b \cos(2\pi * 60 * t + \gamma_b) \quad (5)$$

$$i_c(t) = \sqrt{2}I_c \cos(2\pi * 60 * t + \gamma_c) \quad (6)$$

$\boldsymbol{I}_{dq0}(t)$ is then used to solve the generator rotor differential equations. However, instead of solving the differential equations up to $t + \Delta T$ as would be done in a conventional dynamic simulator, TPDA solves the equations from $t$ to $t + \Delta T - \epsilon$; $\epsilon \ll \Delta T$ (based on simulations run thus far, all $\epsilon$ values smaller than $\Delta T / 10$ give similar results). Since $\Delta T$ is already very small (a minimum value of $1/4^{\text{th}}$ of a cycle is used to capture rotor speed oscillations at twice the fundamental frequency due to unbalance), negligible error is introduced in the dynamic states of the generator rotor from the states obtained if the integration step was $t + \Delta T$.

Next, the new generator states along with $\mathbf{I_{dq0}}(t)$ are used to calculate $\psi_q(t + \Delta T - \epsilon)$ and $\psi_d(t + \Delta T - \epsilon)$ [13] which are substituted in (1)-(3) to obtain $\mathbf{v_{dq0}}(t + \Delta T - \epsilon)$, the 3X1 vector of $dq0$ frame voltages at time $t + \Delta T - \epsilon$. The first three of the desired six $dq0$ frame voltages are now available.

To obtain the remaining three $dq0$ frame voltages, $\mathbf{i_{abc}}(t + \Delta T - \epsilon)$ is calculated using the current phasors obtained at time $t$ since the current waveform does not change between $t$ and $t + \Delta T$. $\mathbf{i_{abc}}(t + \Delta T - \epsilon)$ is transformed into $\mathbf{I_{dq0}}(t + \Delta T - \epsilon)$ using Park's transform and used along with the generator states at $t + \Delta T$ (same as the states at $t + \Delta T - \epsilon$) to obtain $\psi_q(t + \Delta T)$ and $\psi_d(t + \Delta T)$, which when substituted in (1)-(3) gives $\mathbf{v_{dq0}}(t + \Delta T)$. Therefore, six $dq0$ frame voltages are now available to uniquely calculate the three $abc$ frame voltage phasors at the generator terminal.

### B. Calculation of Three Phase Voltage Phasors from Six $dq0$ frame Voltages

Equation (7) shows the general relation between $dq0$ and $abc$ frame voltages [13].

$$\mathbf{v_{dq0}}(t) = \mathbf{P}(t) * \mathbf{v_{abc}}(t) \tag{7}$$

where, $\mathbf{P}(t)$ is the Park Transformation Matrix at time $t$ [13], and $\mathbf{v_{abc}}(t)$ is the vector of $abc$ frame voltages at time $t$.

Assuming that the voltage waveform does not change between $t_1 = t + \Delta T - \epsilon$ and $t_2 = t + \Delta T$, the relation between the $dq0$ and $abc$ frame voltages at $t_1$ and $t_2$ is:

$$\mathbf{v_{dq0}}(t_1) = \mathbf{P}(t_1) * \mathbf{v_{abc}}(t_1) \tag{8}$$
$$\mathbf{v_{dq0}}(t_2) = \mathbf{P}(t_2) * \mathbf{v_{abc}}(t_2) \tag{9}$$

Since $\mathbf{P}(t)$ is always invertible, (8) and (9) can be used to calculate unique values of $\mathbf{v_{abc}}(t_1)$ and $\mathbf{v_{abc}}(t_2)$ using (10) and (11), respectively.

$$\mathbf{v_{abc}}(t_1) = \mathbf{P^{-1}}(t_1) * \mathbf{v_{dq0}}(t_1) \tag{10}$$
$$\mathbf{v_{abc}}(t_2) = \mathbf{P^{-1}}(t_2) * \mathbf{v_{dq0}}(t_2) \tag{11}$$

Once the unique values of $\mathbf{v_{abc}}(t_1)$ and $\mathbf{v_{abc}}(t_2)$ are obtained, the three voltage phasors can be calculated. The derivation for obtaining the phasors is given below.

Let $Ve^{j\theta}$ be the voltage phasor of phase A. Let $x_1$ and $x_2$ be its instantaneous values at $t_1$ and $t_2$ which are equal to the first elements of vectors $\mathbf{v_{abc}}(t_1)$ and $\mathbf{v_{abc}}(t_2)$, respectively.

In the time domain the voltage phasor can be expressed as a cosine waveform such as the one in (12).

$$x(t) = \sqrt{2}V\cos(\omega_s t + \theta) \tag{12}$$

(12) can be expanded using the standard trigonometric identity for the cosine of two angles into:

$$x(t) = \sqrt{2}V\cos(\omega_s t)\cos(\theta) - \sqrt{2}V\sin(\omega_s t)\sin(\theta) \tag{13}$$

Denoting $\sqrt{2}V\cos(\theta)$ by $A$ and $-\sqrt{2}V\sin(\theta)$ by $B$, (13) can be written as:

$$x(t) = A\cos(\omega_s t) + B\sin(\omega_s t) \tag{14}$$

Since $x_1$ and $x_2$ are two samples of (14) at $t_1$ and $t_2$, respectively, we obtain:

$$x_1 = A\cos(\omega_s t_1) + B\sin(\omega_s t_1) \tag{15}$$
$$x_2 = A\cos(\omega_s t_2) + B\sin(\omega_s t_2) \tag{16}$$

$A$ and $B$ can be calculated from (15) and (16) using the formula in (17) as long as $t_2 - t_1 \neq \frac{n\pi}{\omega_s}, n \in \mathbb{Z}_{\geq 0}$.

$$\begin{bmatrix} A \\ B \end{bmatrix} = \frac{\begin{bmatrix} \sin(\omega_s t_2) & -\sin(\omega_s t_1) \\ -\cos(\omega_s t_2) & \cos(\omega_s t_1) \end{bmatrix}}{\sin(\omega_s(t_2 - t_1))} \begin{bmatrix} x_1 \\ x_2 \end{bmatrix} \tag{17}$$

$$\sqrt{2}V\cos(\theta) = \frac{x_1\sin(\omega_s t_2) - x_2\sin(\omega_s t_1)}{\sin(\omega_s(t_2 - t_1))} \tag{18}$$

$$\sqrt{2}V\sin(\theta) = \frac{x_1\cos(\omega_s t_2) - x_2\cos(\omega_s t_1)}{\sin(\omega_s(t_2 - t_1))} \tag{19}$$

The magnitude and phase angle of the voltage waveform can be obtained from (18) and (19) using (20) and (21), respectively.

$$V = \left(\frac{1}{\sqrt{2}}\right)\sqrt{(18)^2 + (19)^2} \tag{20}$$

$$\theta = atan2((19), (18)) \tag{21}$$

In TPDA, phase B and C voltage phasors at time $t + \Delta T$ are also calculated by applying (18) - (21) to the 2nd and 3rd elements of $\mathbf{v_{abc}}(t_1)$ and $\mathbf{v_{abc}}(t_2)$, respectively.

### C. Algorithm used in TPDA to solve the DAEs

The algorithm for simulating electromechanical transients that incorporates the formulation discussed above is presented in the flowchart of Fig. 1; Table II defines the symbols used in Fig. 1. The algorithm shows that TPDA uses a sequential or partitioned method for solving the DAEs [13], [14]; the differential equations are solved using the trapezoidal method as implemented in the ode23t function of MATLAB while the Distributed Engineering Workstation (DEW®) software [16] is used to solve the algebraic equations. The four reasons for selecting this approach are as follows:

1. Conceptual and implementation simplicity; algebraic and differential equation solvers can be selected independently.

2. Differential equations can be solved in any order and in parallel over multiple processors.

3. While any nonlinear equation solver can be used to solve the network algebraic equations, TPDA uses DEW® because the algorithm used in DEW® can be easily modified to split a network into multiple radial sections which can be solved in parallel across multiple processors, thereby significantly reducing the simulation time.

4. DEW® has been used to model real utility networks that contain more than 3 million components (lines, switches, loads, etc.), and the only limitation encountered has been the physical memory of the machine. Therefore, using DEW to solve the network algebraic equations provides TPDA the capability to simulate electromechanical transients in very large networks.

### III. Validation of TPDA & its Application for Studying Electromechanical transients in Hybrid Models

Validation of TPDA with the WECC 9 bus system was performed in [17]. In this section of the paper three case studies are discussed which are designed to achieve the following objectives:

1. Demonstrate the accuracy of TPDA in simulating electromechanical transients under balanced and unbalanced network conditions by comparing simulation results with GE-PSLF® and the Alternative Transients Program (ATP), respectively (case studies 1 and 2).

2. Demonstrate the ability of TPDA to simulate electromechanical transients in large, real utility, hybrid models (case study 3).

3. Highlight the advantages of studying electromechanical transients using hybrid models (case study 3).

The case studies are described in Tables III & IV. For case studies 1 and 2, rotor speed deviation (rotor speed minus synchronous speed) and terminal voltages obtained using TPDA are compared with calculations from GE-PSLF® and ATP. The following measures are used to present this comparison:

1. **Plots of trajectories**: to provide a visual representation of the accuracy of TPDA.
2. **Correlation coefficients**: to quantify <u>degree of match in the shape</u> of the trajectories of the comparison variables.
3. **Root Mean Square Errors** (RMSEs): to quantify the <u>degree of match in actual value</u> of the comparison variables.

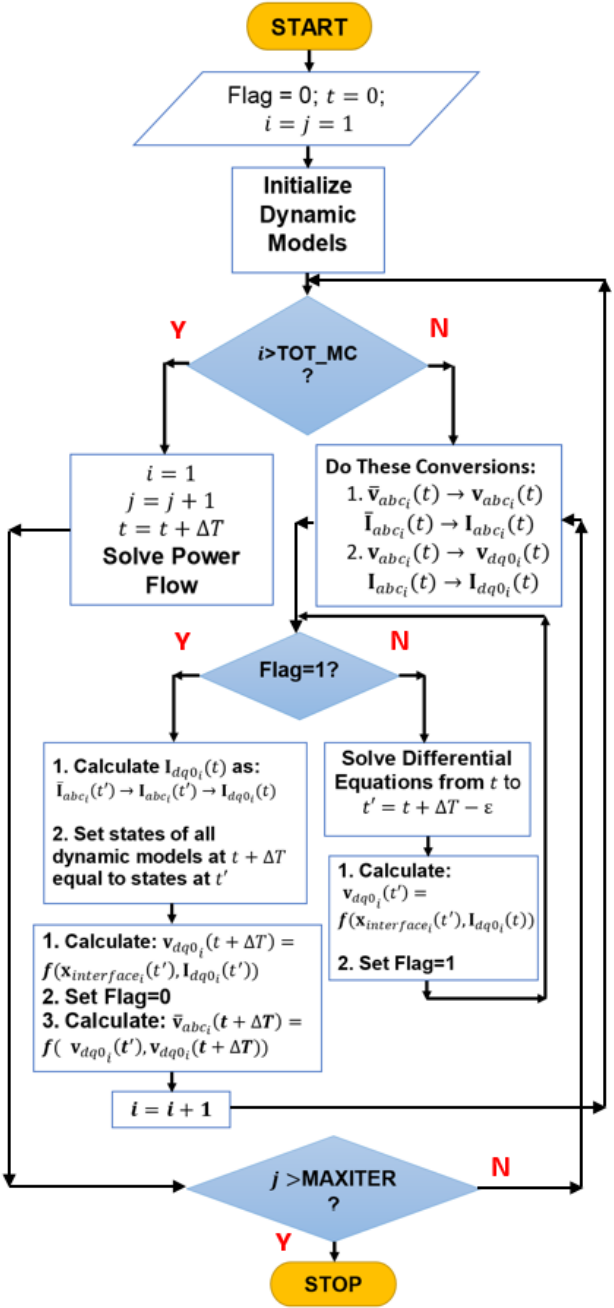

Fig. 1. Flowchart of algorithm

TABLE II
Definition of Symbols Used in Flowchart of Fig. 1

| Symbol | Definition |
|---|---|
| $i; j$ | Counter for buses with dynamic models; counter for simulation iterations |
| TOT_MC; MAXITER | Constant representing total # of buses with dynamic models; Constant representing total # of simulation iterations |
| $\Delta T; \epsilon$ | Integration time step; a small number $<<\Delta T$ |
| $\bar{\mathbf{v}}_{abc_i}(t); \bar{\mathbf{I}}_{abc_i}(t)$ | Vectors of voltage and current phasor at bus $i$ at time $t$ |
| $\mathbf{v}_{abc_i}(t); \mathbf{I}_{abc_i}(t)$ | Instantaneous voltage and current vectors for bus $i$ at time $t$ |
| $\mathbf{v}_{dq0_i}(t); \mathbf{I}_{dq0_i}(t)$ | $dq0$ frame voltage and current vectors for bus $i$ at time $t$ |
| Flag | Ensures that at each simulation iteration $\mathbf{v}_{dq0_i}(t + \Delta T - \epsilon)$ and $\mathbf{v}_{dq0_i}(t + \Delta T)$ are correctly calculated |
| $\mathbf{x}_{interface_i}(t)$ | State vector of dynamic model (e.g. synchronous generator) that directly connects at the $i^{th}$ bus |

TABLE III
Description of Case Studies

| Case Study # | Network Topology (Table IV) | Dynamic Models | Disturbance |
|---|---|---|---|
| 1 | IEEE 39 Bus | Generator (GENROU) | 1,500 MW; 552 MVAR balanced increase at Bus 4 (3X original load) * |
| 2 | IEEE 39 Bus | Generator (GENROU) | 280 MW; 2,870 MVAR increase on Phase A at Bus 12 (99X original load) * |
| 3 | Utility Model | Generator (GENROU); Substation (Infinite Bus) | Phase A to ground fault at a 60 kV substation; 0.2 ohm fault impedance |

* For case studies 1&2 disturbance was initiated at 0.1 & removed at 0.3 second

TABLE IV
Description of Network Topology

| Component Type | IEEE 39 Bus | Utility Model |
|---|---|---|
| 2-Phase Lines/Cables | 0 | 1,758 |
| 3-Phase Lines/Cables | 34 | 7,472 |
| 1-Phase Transformers | 0 | 2,655 |
| 3-Phase Transformers | 12 | 1,410 |
| Fixed Shunt Capacitors | 0 | 62 |
| Switched Shunt Capacitors | 0 | 38 |
| Breakers and Switches | 0 | 7,526 |
| Total Loads (3-Phase, 2-Phase and 1-Phase) | 19* | 4,153** |
| **Total Elements** | **65** | **25,074** |

* Only 3-Phase loads; Constant impedance load model is assumed
** ZIP model for load on each phase

Due to limited space, results are provided for selected buses only. Buses are selected such that results from multiple locations in the networks can be presented. The parameters of dynamic models used in case studies 1 and 2 are given in [18].

### A. Case Study 1: Validation of TPDA with PSLF

The network topology used in case studies 1 and 2 is shown in Fig. 2 [19]. The buses where disturbances were simulated and for which results are provided in the figures and tables that follow are indicated in Fig. 2.

Fig. 3 and 4 and Table V show that the trajectories generated by PSLF and TPDA match very well and the RMSEs for all the buses are very small while the correlation coefficients are close to unity.

### B. Case Study 2: Validation of TPDA with ATP

Similar to case study 1, rotor speed deviation and terminal voltages calculated by TPDA and ATP match very well as seen in Fig. 5 and 6. Moreover, Tables VI and VII show that the RMSEs are small and correlation coefficients are close to 1.

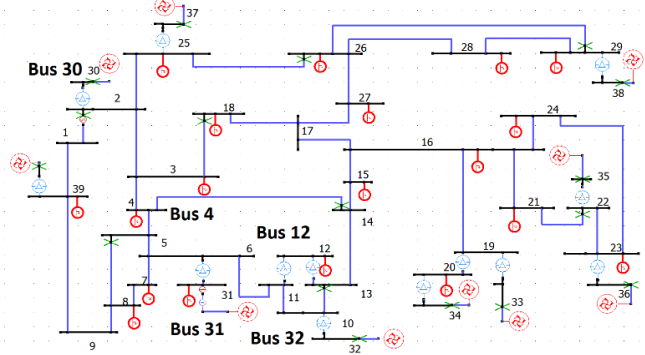

Fig. 2. IEEE 39 bus system

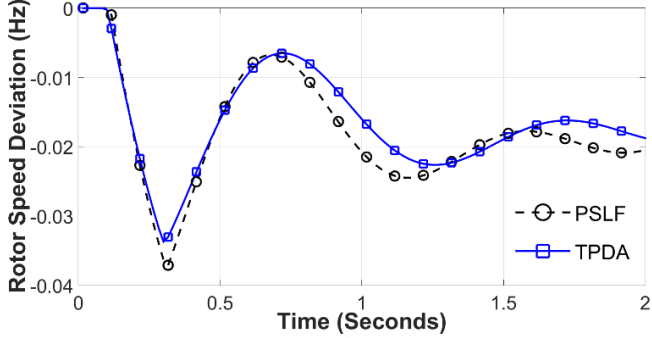

Fig. 3. Case study 1: rotor speed deviation of generator at bus 30

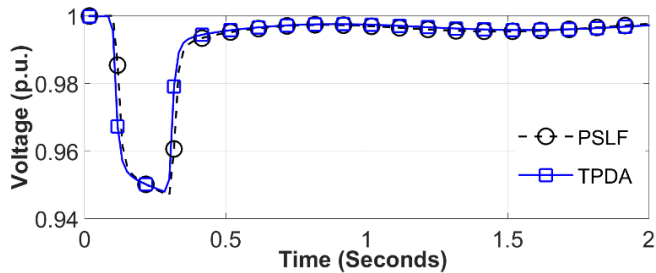

Fig. 4. Case study 1: terminal voltage of generator at bus 31

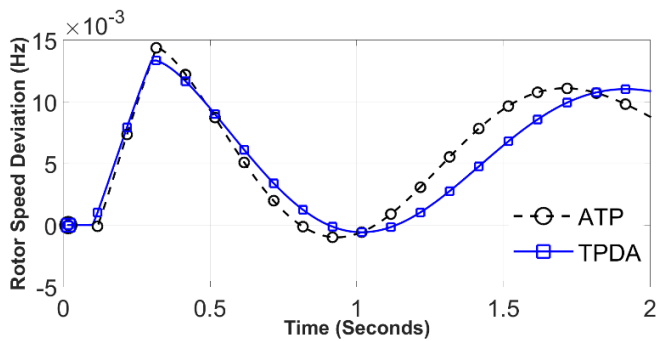

Fig. 5. Case study 2: rotor speed deviation of generator at Bus 32

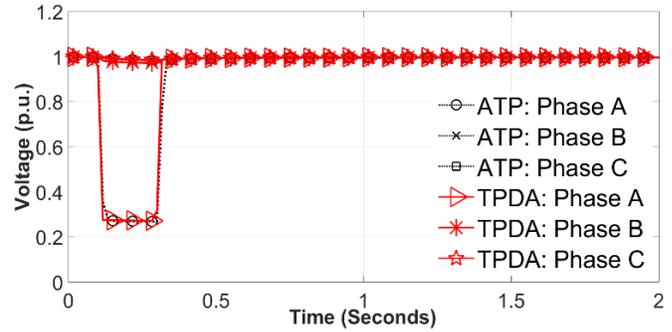

Fig. 6. Case study 2: three phase voltage at bus 12

TABLE V
Case Study 1: Correlation Coefficients and RMSE between TPDA and PSLF

| Generator Bus # | Correlation Coefficients | | Root Mean Square Error | |
|---|---|---|---|---|
| | Rotor Speed | Terminal Voltage | Rotor Speed (Hz) | Terminal Voltage (p.u.) |
| 30 | 0.97 | 0.99 | 0.002 | 0.0004 |
| 31 | 0.92 | 0.98 | 0.005 | 0.0026 |
| 32 | 0.96 | 0.94 | 0.006 | 0.0007 |
| 33 | 0.98 | 0.99 | 0.002 | 0.0003 |
| 34 | 0.99 | 0.81 | 0.002 | 0.0003 |
| 35 | 0.99 | 0.99 | 0.002 | 0.0004 |
| 36 | 0.98 | 0.95 | 0.002 | 0.0004 |
| 37 | 0.92 | 0.98 | 0.003 | 0.0004 |
| 38 | 0.97 | 0.96 | 0.002 | 0.0003 |
| 39 | 0.98 | 0.99 | 0.002 | 0.0003 |

TABLE VI
Case Study 2: Correlation Coefficients and RMSE of Rotor Speed Deviation between TPDA and ATP

| Generator Bus # | Correlation Coefficients | Root Mean Square Error (Hz) |
|---|---|---|
| 30 | 0.99 | 0.0003 |
| 31 | 0.98 | 0.0013 |
| 32 | 0.95 | 0.0015 |
| 33 | 0.98 | 0.0004 |
| 34 | 0.98 | 0.0005 |
| 35 | 0.99 | 0.0003 |
| 36 | 0.99 | 0.0003 |
| 37 | 0.98 | 0.0004 |
| 38 | 0.97 | 0.0007 |
| 39 | 0.99 | 0.0004 |

TABLE VII
Case Study 2: Correlation Coefficients and RMSE of Bus 12 Voltage between TPDA and ATP

| | Correlation Coefficients | Root Mean Square Error (p.u.) |
|---|---|---|
| Phase A | 0.99 | 0.0227 |
| Phase B | 0.92 | 0.0042 |
| Phase C | 1.00 | 0.0011 |

## C. Case Study 3: Transient Simulation with the Utility Model

### 1): Description of the Case Study

The Utility Model is the three phase hybrid model of a North American utility. The basic configuration of the network is shown in Fig. 7. Total load modeled in the case study is 370 MW. 109 MW of this load is served from a power plant owned by the utility and remaining 261 MW is imported from the neighboring utility through a 230 kV substation (149 MW) and multiple 115 kV tie lines (112 MW). GENROU is the dynamic model used for the utility owned power plant, the substation is modeled as an infinite bus, and the tie line flows are modeled as constant power injections.

To highlight the advantages of using hybrid models, the case study is simulated in two parts. First, the hybrid model with detailed distribution feeder models is simulated. Next, the "Transmission only Model (T-model)" is used in which all the distribution feeders are de-energized and their loads are lumped together at the corresponding 60 kV buses as constant power loads. Fig. 8 shows the configuration of a substation in the hybrid model (Fig. 8c) and the T-model (Fig 8a).

### 2): Description of Fault Simulation and Fault Clearing

For both the T- model and the hybrid model the single line to ground (SLG) fault is assumed to occur on phase A of a 60 kV substation, just to the right of breaker B2 of Fig. 8a and Fig. 8c. This substation is serving 46 MW before the fault where Feeder 1 (or S1) serves 14 MW (3% load imbalance) and Feeder 2 (or S2) serves 22 MW (0.3% load imbalance). The fault is initiated at the end of the 50th cycle and cleared at the end of the 60th cycle by opening breakers B1 and B2 as shown in Fig. 8b and 8d. Clearing the fault results in loss of power to load S2 in the T-model (Fig. 8b) and Feeder 2 in the hybrid model (Fig. 8d). However, since the detailed substation configuration of the 12kV distribution network is included in the hybrid model, the normally open breaker B3 is closed 30 cycles after the fault is cleared to restore power to Feeder 2 (Fig. 8d).

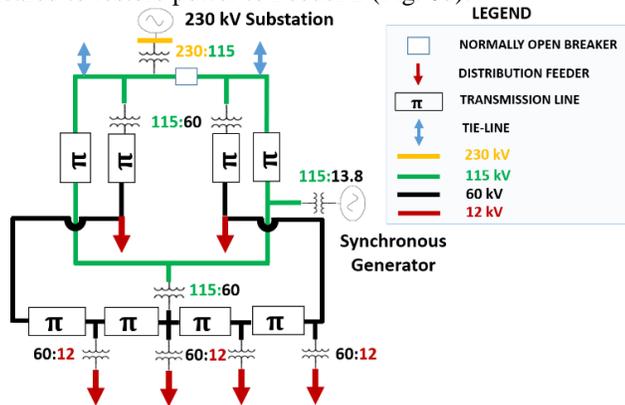

Fig. 7. Configuration of the Utility Model

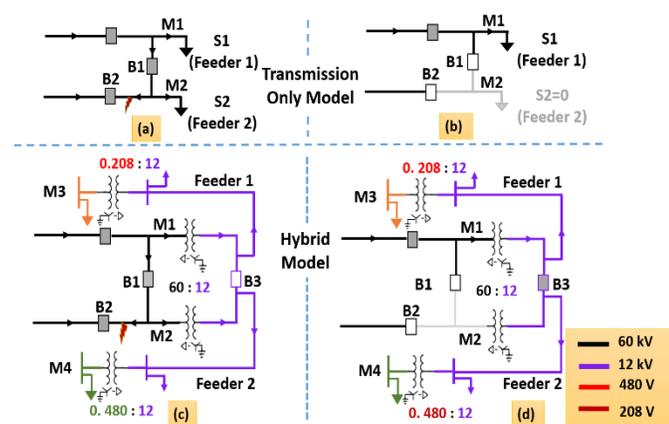

Fig. 8. Pre-fault, during fault and post-fault substation configuration

### 3): Important Observations from the Case Study

Simulation of electromechanical transients in hybrid models using TPDA can provide utility engineers with insights that cannot be obtained from transmission only models. Two such insights were obtained from this case study − (i) the impact of voltage sags on customers; and (ii) the ability of the network to restore power to customers by reconfiguration. These are discussed in detail below.

#### a): Voltage sags due to the fault

Voltage sags are an important power quality issue for the electric power industry and about 70% of them are caused by SLG faults [20]. The ITI curve [20] is an industry standard voltage vs. duration curve that was primarily developed to identify safe, prohibited and reduced/abnormal performance regions of 120 volt information technology (IT) equipment [http://www.powerqualityworld.com/2011/04/itic-power-acceptability-curve.html]. Similar curves for residential equipment such as air-conditioners, microwave ovens, televisions, etc. were developed in [20]. Both these curves are used in the discussion that follows. This discussion is based on Fig. 9 and Fig. 10 which show the voltage trajectories at the 60 kV bus of the T-model (M1; Fig. 8a) and the 120 V (line-ground) bus of the hybrid model (M3; Fig. 8c) that is supplied through M1.

- **Phase A Voltage Sag during Fault (Fig. 9 & Fig. 10):**
  Phase A voltage at M1 in the T-model drops to and stays below 0.5 p.u. during the fault, thereby moving into the region of the ITI curve where IT equipment may perform abnormally (Fig. 10). According to [20], computers are also likely to restart. However, phase A voltage at M3 remains above 0.85 p.u. Voltages observed at several distribution buses (not shown here) were also in the "no effect of sag" region of the ITI curve shown in Fig. 10.

  Therefore, while the T-model suggests that ITI compliant equipment connected to Phase A and residential devices such as computers and microwave ovens [20] are likely to function abnormally, the more accurate hybrid model shows that none of these power quality issues will occur as a result of the fault.

- **Phase B & C Voltages during Fault (Fig. 9):** The T-only model shows that phase B and C voltages at M1 *increase* during the fault, although they are within the normal operating region of the ITI curve. On the contrary, the hybrid model shows that phase B and C voltages at M3 *reduce* significantly from their pre-fault values with phase B voltage staying inside the "effect of sag" region of the ITI curve.

These observations demonstrate that studying electromechanical transients in hybrid models using TPDA can prevent engineers from making erroneous power quality assessments that could be made with transmission only models.

#### b): Reconfiguration to restore power to Feeder 2

As mentioned earlier, the hybrid model makes it possible to restore power in Feeder 2 (Fig. 8d) which is lost when the fault is cleared. Fig. 11 shows that after the fault is cleared voltages at M2 in the T-model stay at zero throughout the simulation while they return to the pre-fault steady state at M4 (Fig. 8d), which is served by Feeder 2 in the hybrid model.

Moreover, as shown in Fig. 12, the generator rotor speed also returns to the pre-fault steady state value after load restoration indicating that the network has returned to a stable operating point.

These observations demonstrate that TPDA, when used in conjunction with hybrid models, not only allows utility engineers to evaluate various reconfiguration scenarios for power restoration, but also identify the scenarios that will result in stable operation of the network.

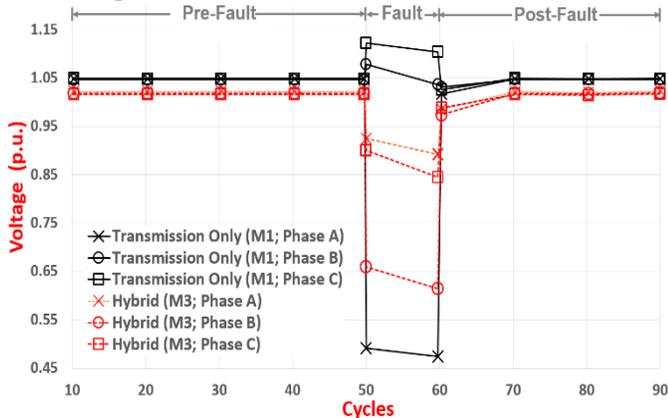

Fig. 9. Voltage at M1 in the T-model and M3 in the hybrid model of Fig. 8

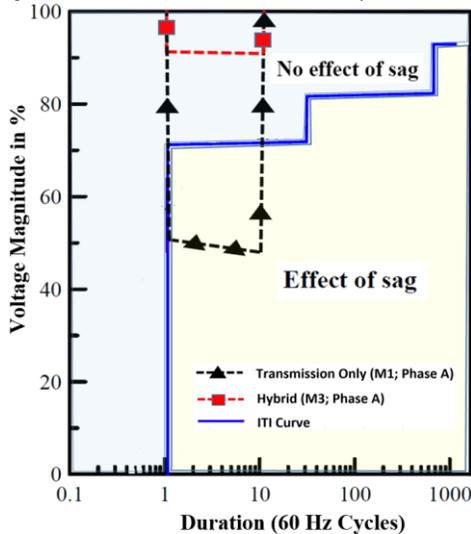

Fig. 10. Phase A voltage at M1 in the T-model and M3 in the hybrid model of Fig. 8 superimposed on the under-voltage region of the ITI curve which is taken from [20]

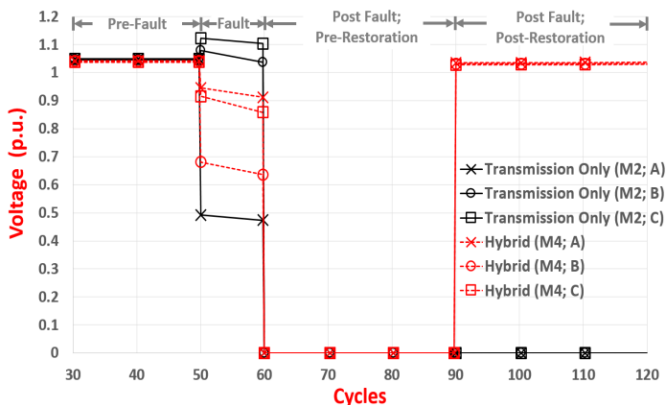

Fig. 11. Voltage at M2 in the T-model and M4 in the hybrid model of Fig. 8

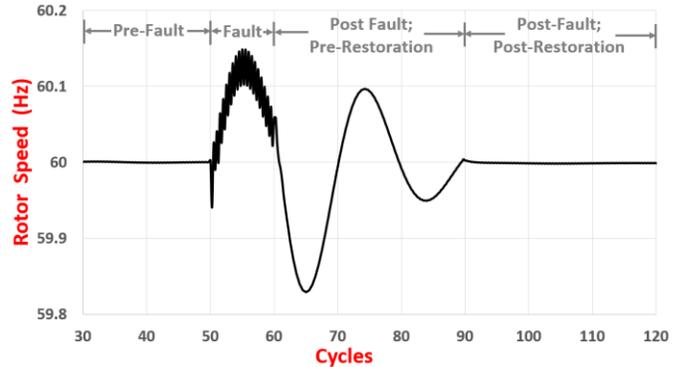

Fig. 12. Rotor speed of the synchronous generator in the hybrid model

## IV. ADVANTAGES OF TPDA OVER EMTP PROGRAMS; SCOPE AND LIMITATIONS

### 1): Scope and Limitations of TPDA

TPDA models the electric network using algebraic equations and assumes that the network transitions from one state to the other instantaneously at the fundamental frequency. Therefore, TPDA can be used for studying electromechanical transients using three phase network models. However, because of these assumptions, it cannot be used to study electromagnetic transients.

### 2): Advantages of TPDA over EMTP Programs

EMTP programs are indispensable if electromagnetic transients are to be studied. However, if the objective is to study electromechanical transients using network models that accurately model un-transposed lines, unbalanced loads, detailed distribution feeders, asymmetrical faults, or roof top solar PV transients, then TPDA has several advantages over EMTP programs. These include:

1. **Computation Time**: case study 3 used a hybrid model of an actual utility which contained more than 25,000 elements. TPDA simulated 1 second of this model in 16 minutes using an integration step of 4 samples/cycle. [21] reports that when EMTP-RV was used to simulate a single line to ground fault in a distribution network that contained around 37,000 elements, it took 9.7 hours to solve the main system of equations for 11 cycles implying that a full second of simulation would have taken about 53 hours − 134 times longer than TPDA after adjusting for different system sizes. The machine used in case study 3 had 8 GB RAM and 2.4 GHz Intel Core i7-5500U processor while that used in [21] had 24 GB RAM and 3.33 GHz Intel Core i7-975 processor.

The step size used in [21] was 256 samples/cycle. To the best of our knowledge the network used in [21] did not model synchronous machines or the transmission network. Including these models increases the stiffness of differential equations, which may require smaller integration time steps, thereby increasing the simulation time. For example, in case studies 1 and 2 it was observed that step size smaller than 245 samples/cycle caused ATP to become numerically unstable, while TPDA converged with a step size of 4 samples/cycle.

2. **Network Size**: According to the ATP website (http://www.emtp.org/), the maximum number of branches that

can be modeled in the standard EEUG program distribution is 10,000. As shown in Table IV, the Utility Model of case study 3 contains over 13,000 branches. Similarly, the maximum number of switches that can be included in ATP are 1200, while TPDA models 7,526 switches and breakers in case study 3.

Since TPDA uses DEW® for solving the algebraic equations, system size is not a limitation. Actual utility networks with more than 3 million components have been modeled in DEW® on 32 bit desktop machines.

**3. Convenient and Economic Deployment**: TPDA uses the same network models that are used by steady state analysis applications in DEW®. Therefore, additional expenditure and inconvenience of deploying and maintaining a separate EMTP program and migrating data and models is avoided.

## V. Conclusion

This paper introduces a new method and software tool, TPDA, for studying electromechanical transients using three phase network models. Through three case studies it is shown that TPDA can accurately simulate electromechanical transients under balanced and unbalanced network conditions, and reveals useful engineering information from the simulation of hybrid models that cannot be obtained from transmission only models.

Our next objective is to use TPDA to study the impact of DG, particularly solar PV, on the stability of power systems using hybrid models of utilities. We are working on developing the three phase and single phase DG dynamic models that are needed for the study. We hope to share the results of this effort with the power systems community in the near future.

## VI. Acknowledgment

The authors gratefully acknowledge the guidance provided by Dr. Steve Southward at Virginia Tech during the development of the methodology discussed in this paper.

## VII. References

[1] U.S. Energy Information Administration, "Energy Outlook 2014," U.S. Department of Energy, Washington, DC, DOE/EIA-0383(2014), Apr. 2014. [Online]. Available: http://www.eia.gov/forecasts/archive/aeo14/pdf/0383(2014).pdf

[2] K. Kroh, "California Installed More Rooftop Solar in 2013 Than Previous 30 Years Combined", Think Progress, Jan. 2014. [Online]. Available: http://thinkprogress.org/climate/2014/01/02/3110731/california-rooftop-solar-2013/

[3] I. Xyngi, A. Ischenko, M. Popov, and L. Sluis, "Transient Stability Analysis of a Distribution Network with Distributed Generators," *IEEE Trans. Power Systems*, vol. 24, pp. 1102-1104, May 2009.

[4] R.S. Thallam, S. Suryanarayanan, G.T. Heydt, and R. Ayyanar, "Impact of Interconnection of Distributed Generation on Electric Distribution Systems – A Dynamic Simulation Perspective," in *Proc. 2006 IEEE Power Engineering Society General Meeting*.

[5] Z. Miao, M.A. Choudhry, and R.L. Klein, "Dynamic simulation and stability control of three-phase power distribution system with distributed generators," in *Proc. 2002 IEEE Power Engineering Society Winter Meeting*, pp.1029 – 1035.

[6] B.W. Lee and S.B. Rhee, "Test Requirements and Performance Evaluation for Both Resistive and Inductive Superconducting Fault Current Limiters for 22.9 kV Electric Distribution Network in Korea," *IEEE Trans. Applied Superconductivity*, vol. 20, pp. 1114-1117, June 2010.

[7] E.N. Azadani, C. Canizares, C and K. Bhattacharya, "Modeling and stability analysis of distributed generation," in *Proc. 2002 IEEE Power and Energy Society General Meeting*.

[8] E. Nasr-Azadani, C.A. Canizares, D.E. Olivares, and K. Bhattacharya, "Stability Analysis of Unbalanced Distribution Systems With Synchronous Machine and DFIG Based Distributed Generators," *IEEE Trans. Smart Grid*, vol.5, pp. 2326-2338, Sept. 2014.

[9] Xuefeng Bai, Tong Jiang, Zhizhong Guo, Zheng Yan and Yixin Ni, "A unified approach for processing unbalanced conditions in transient stability calculations," *IEEE Trans. Power Systems*, vol.21, pp. 85-90, Feb. 2006.

[10] M. Reza, J.G. Slootweg, P.H. Schavemaker, W.L. Kling, L van der Sluis., "Investigating impacts of distributed generation on transmission system stability," in *Proc. 2003 IEEE Bologna Power Tech Conference*.

[11] Black & Veatch, "Biennial Report on Impacts of Distributed Generation," California Public Utilities Commission, B&V Project No. - 176365, May 2013. [Online]. Available: http://www.cpuc.ca.gov/NR/rdonlyres/BE24C491-6B27-400C-A174-85F9B67F8C9B/0/CPUCDGImpactReportFinal2013_05_23.pdf

[12] R.H. Salim and R.A. Ramos, "A Model-Based Approach for Small-Signal Stability Assessment of Unbalanced Power Systems," *IEEE Trans. Power Systems*, vol.27, pp.2006-2014, Nov. 2012.

[13] P.W. Sauer and M.A. Pai, *Power System Dynamics and Stability*, Illinois: Stipes Publishing L.L.C, 2006, p. 26, 35, 47, 155, 165, 195.

[14] P. Kundur, *Power System Stability and Control*, New Delhi: Tata McGraw Hill, 2012, p. 858, 861.

[15] S. Abhyankar, "Development of an Implicitly Coupled Electromechanical and Electromagnetic Transients Simulator for Power Systems," Ph.D. dissertation, Dept. Elect. Eng., Illinois Institute of Technology, Chicago, 2011.

[16] M. Dilek, F. de Leon, R. Broadwater, and S. Lee, "A Robust Multiphase Power Flow for General Distribution Networks", *IEEE Trans. Power Systems*, vol. 25, pp. 760-768, May 2010.

[17] H. Jain, A. Parchure, R.P. Broadwater, M. Dilek, J. Woyak., "Three Phase Dynamics Analyzer - A New Program for Dynamic Simulation using Three Phase Models of Power Systems," in *Proc. 2015 IEEE IAS Joint ICPS/PCIC Conference*.

[18] Parameters for Dynamic Models of Power Plant Equipment. [Online]. Available: https://www.scribd.com/doc/283485862/Parameters-for-Dynamic-Models-of-Power-Plant-Equipment

[19] IEEE 39 Bus System, Illinois Center for a Smarter Electric Grid (ICSEG). [Online]. Available: http://publish.illinois.edu/smartergrid/ieee-39-bus-system/

[20] G.G. Karady, S. Saksena, B. Shi, N. Senroy, "Effects of Voltage Sags on Loads in a Distribution System," Power Systems Energy Research Center, Ithaca, NY, PSERC Publication 05-63, Oct. 2005.

[21] V. Spitsa, R. Salcedo, Ran Xuanchang, J.F. Martinez, R.E. Uosef, F. de Leon, D. Czarkowski and Z. Zabar, "Three–Phase Time–Domain Simulation of Very Large Distribution Networks," *IEEE Trans. Power Delivery*, vol.27, pp.677-687, April 2012.

## VIII. Biographies

**Himanshu Jain** is pursuing his PhD in Electrical Engineering at Virginia Tech. He holds a MS degree in Electrical Engineering from the University of Texas at Arlington and a B.Tech degree in Electrical Engineering from G.B. Pant University of Agriculture and Technology. His research interests include power system dynamics, renewable integration, and electricity markets.

**Abhineet Parchure** is currently pursuing his MS in Electrical Engineering at Virginia Tech. He received a B.E. (Hons.) degree in Electrical and Electronics Engineering from Birla Institute of Technology & Science, Pilani in 2012. His research interests include renewable energy integration, power system dyanmics and voltage stability of transmission and distribution systems.

**Robert P. Broadwater** (M'71) received the B.S., M.S., and Ph.D. degrees in electrical engineering from Virginia Polytechnic Institute and State University (Virginia Tech), Blacksburg, VA, in 1971, 1974, and 1977, respectively. He is currently a Professor of electrical engineering at Virginia Tech. He develops software for analysis, design, operation, and real-time control of physical systems. His research interests are object-oriented analysis and design and computer-aided engineering.

**Murat Dilek** received the M.S. and Ph.D. degrees in electrical engineering from Virginia Tech, Blacksburg, in 1996 and 2001, respectively. He is a Senior

Development Engineer at Electrical Distribution Design, Inc. His work involves computer-aided design and analysis of electrical power systems.

**Jeremy Woyak** (S'07) received a B.S. degree in electrical engineering from Lawrence Technological University, Southfield, MI, in 2010 and a M.S. degree in electrical engineering from Virginia Polytechnic Institute and State University, Blacksburg, VA, in 2012. He is currently a software developer at Electrical Distribution Design. He develops software for analysis, design, operation, and real-time control of electrical power systems.